\documentclass[12pt,titlepage]{article}
\usepackage{amsmath}
\usepackage{amssymb}
\usepackage{amsthm}
\usepackage{booktabs}
\usepackage{graphicx}
\usepackage{subfigure}
\usepackage[round]{natbib}
\newtheorem{proposition}{Proposition}
\newtheorem{corollary}{Corollary}

\newcommand\indep{\protect\mathpalette{\protect\independenT}{\perp}}
\def\independenT#1#2{\mathrel{\rlap{$#1#2$}\mkern2mu{#1#2}}}

\begin{document}
\title{Instrumental Variable Estimation When Compliance is not Deterministic: The Stochastic Monotonicity Assumption}
\thanks{Address for Correspondence: Dylan Small, Department of Statistics, The Wharton School of the University of Pennsylvania, Philadelphia, PA 19104 (E-mail: {\it{dsmall@wharton.upenn.edu}}).}
\author{Dylan S. Small\\University of Pennsylvania \and Zhiqiang Tan\\Rutgers University \and Scott A. Lorch\\University of Pennsylvania \and M. Alan Brookhart\\University of North Carolina}
\date{\today}
\maketitle

\pagenumbering{gobble}
\clearpage
\thispagestyle{empty}

{\bf{Abstract}}: The instrumental variables (IV) method is a method
for making causal inferences about the effect of a treatment based on
an observational study in which there are unmeasured confounding
variables.  The method requires a valid IV, a variable that is
independent of the unmeasured confounding variables and is associated
with the treatment but which has no effect on the outcome beyond its
effect on the treatment.  An additional assumption that is often made
for the IV method is deterministic monotonicity, which is an
assumption that for each subject, the level of the treatment that a
subject would take if given a level of the IV is a monotonic
increasing function of the level of the IV.  Under deterministic
monotonicity, the IV method identifies the average treatment effect for the compliers (those subject who would take the treatment if encouraged to do so by the IV and not take the treatment if not encouraged).  However, deterministic monotonicity is sometimes not realistic.
We introduce a stochastic monotonicity condition which relaxes deterministic monotonicity in that it does not require that a
monotonic increasing relationship hold within subjects between the levels of the IV and
the level of the treatment that the subject would take if given a
level of the IV, but only that a monotonic
increasing relationship hold across subjects between the IV and the treatment in a
certain manner.  We show that under stochastic monotonicity, the IV method identifies a weighted average of treatment effects with greater weight on subgroups of subjects on whom the IV has a stronger effect.  We provide bounds on the global average treatment effect under stochastic monotonicity and a sensitivity analysis for violations of the stochastic monotonicity assumption.  We apply the methods of IV inference under stochastic monotonicity that we develop to a study of the effect of premature babies being delivered in a high volume, high technology neonatal intensive care unit (NICU) vs. a lower level unit.


\newpage

\clearpage
\pagenumbering{arabic}

\bf 1. Introduction \rm

The instrumental variable (IV) method is a method for estimating the effect of a treatment when there is unmeasured confounding. \cite{angrist1996identification} presented a framework for IV analysis that illuminated the causal effect that is estimated by an IV analysis for many settings.  \cite{angrist1996identification} defined a binary IV as a variable that (1) has two levels, one that encourages taking treatment and another that does not encourage taking treatment, and the level that encourages taking treatment has a causal effect on taking the treatment; (2) satisfies the deterministic monotonicity assumption that a subject's level of treatment is always at least as high for the level of the IV that encourages a subject to take a treatment compared to the level that does not encourage taking the treatment; (3) is independent of unmeasured confounders; (4) only affects the outcome through affecting whether the subject takes the treatment (the exclusion restriction); and (5) satisfies the stable unit treatment value assumption, which says that there is no interference between units and there are not different versions of the IV nor the treatment.  \citet{angrist1996identification} divide subjects into four compliance classes that describe a subject's behavior under the encouragement and no encouragement level of the IV: (i) always takers, subjects who will take the treatment whether or not they are encouraged to do so by the IV; (ii) never takers, subjects who will not take the treatment regardless of whether or not they are encouraged to do so by the IV ;(iii) compliers, subjects who will take the treatment if encouraged to do so by the IV but not take the treatment if not encouraged by the IV and (iv) defiers, subjects who will not take the treatment if encouraged by the IV but will take the treatment if not encouraged. The deterministic monotonicity assumption says that there are no defiers.  Under the IV assumptions (1)-(5) described above, \citet{angrist1996identification} showed that the Wald estimate, which is the difference in the mean outcome between subjects with the encouraging level of the IV vs. the not encouraging level of the IV divided by the difference in the proportion of subjects taking the treatment with the encouraging level of the IV vs. the not encouraging level of the IV, converges to the average treatment effect for the compliers, called the local average treatment effect (LATE). The average treatment effect for never takers is not identified because these subjects are only ever observed taking level $0$ of the treatment; likewise, the average treatment effects for always takers is not identified because these subjects are only ever observed taking level $1$ of the treatment.

\citet{angrist1996identification}'s framework is ideally suited for a setting where the IV is delivered in a uniform way.  For example,
\citet{finkelstein2012} studied the effect of having health insurance vs. not having health insurance on health care use, financial strain, and the health of low-income adults.  In 2008, the state of Oregon opened a waiting list for a limited number of spots in its Medicaid program for low-income adults, which had previously been closed to new enrollment.  Because the waiting list exceeded the number of spots available, the state drew names by lottery to decide who would have the opportunity to enroll in Medicaid.  Not all winners enrolled in Medicaid either because they did not apply or were deemed ineligible, and some non-winners obtained health insurance through other means;  however, winners had about a 25 percentage point higher chance of having health insurance one year after the lottery.  \citet{finkelstein2012} used winning the lottery vs. not winning as an IV for estimating the effect of health insurance.  Deterministic monotonicity is plausible because winning the lottery provides an additional opportunity to obtain health insurance by enrolling in Medicaid and no disincentive to obtain health insurance.

For an IV that is delivered in a uniform way like the Oregon Medicaid lottery, a subject's compliance class (always taker, never taker, complier or defier) is well defined.  However, many IVs used in observational studies are not delivered in a uniform way and a subject's compliance class is ill defined -- a subject may be an always taker if the IV is delivered in one way but a complier if the IV is delivered in another way \citep{swanson2014b}.  The following are several examples where the IV is not delivered in a uniform way:
\begin{itemize}
\item {\it{Physician prescribing preference IV}}.  For comparing the effectiveness or safety of two drugs, physician's prescribing preference has often been used as an IV. For example, \citet{brookhart2006} compared the benefits and risks of two types of nonsteroidal anti-inflammatory drugs (NSAIDs) for treating gastrointestinal (GI) problems, COX-2 inhibitors versus nonselective nonsteroidal anti-inflammatory drugs (nonselective NSAIDs).  \citet{brookhart2006} used whether the patient's physician's most recent NSAID prescription (prior to the current patient) was a COX-2 inhibitor as an IV for whether the current patient's prescription is a COX-2 inhibitor.  A patient's compliance class depends not only on the patient's physician's preferences but also on the somewhat random factor of what type of patient the physician last saw and is in this sense not well defined.  For example, consider Table 1 which shows the drug types that two hypothetical doctors, Dr. X and Dr. Y, would give to four hypothetical patients, Alice, Bob, Carol and Dave.  Suppose that Dave is the current patient and Alice, Bob and Carol preceded Dave in time.  If Carol was Dr. X's most recent patient and Alice was Dr. Y's most recent patient, then Dave is a complier.  However, if Alice was Dr. X's most recent patient and Carol was Dr. Y's most recent patient, then Dave is a defier.
\begin{table}[h!]
\begin{center}
\begin{tabular}{|c|c|c|}
\hline Patient & Dr. X & Dr. Y \\ \hline
Alice & Nonselective NSAID & Nonselective NSAID \\
Bob & COX-2 inhibitor & COX-2 inhibitor \\
Carol & COX-2 inhibitor & COX-2 inhibitor \\ \hline
Dave & COX-2 inhibitor & Nonselective NSAID \\
\hline
\end{tabular}
\caption{Type of drug that would be given by two hypothetical doctors, Dr. X and Dr. Y, to four hypothetical patients.}
\end{center}
\label{hypothetical.patient.table}
\end{table}

\item {\it{Geographic or Hospital Preference-Based IVs}}.  A similar issue as with the physician prescribing preference IV arises for other preference-based IVs.  A preference-based IV is an IV based on different providers or group of providers (e.g., hospitals or geographic regions) having different preferences or treatment algorithms dictating how medications or medical procedures are used \citep{brookhart2007}.  For example, \citet{brooks2003} studied the effect of mastectomy versus surgery plus irradiation for breast cancer patients using whether the patient's geographic region had a high rate of mastectomy as the IV.   Suppose Region $A$ has a high rate of mastectomy and region $B$ does not.  A patient's compliance class will depend on the specific doctor the patient would go to if living in region $A$ vs. region $B$.  A patient might be a complier if the doctor the patient would go to if living in region $A$ has typical preferences for a doctor in region $A$ and the doctor the patient would go to if living in region $B$ has typical preferences for region $B$.  But in region $A$ there might be some atypical doctors who prefer surgery plus irradiation and in region $B$, there might be some atypical doctors who prefer mastectomy.  If the doctor the patient would go to in region $A$ is atypical and the doctor the patient would go to in region $B$ is atypical, then the patient will be a defier.
\item {\it{Distance to specialty care provider IV}}.  When comparing two treatments, one of which is only provided by specialty care providers and one of which is provided by more general providers, whether a person lives near or far from the specialty care provider has often been used as an IV, e.g., \citet{mcclellan1994does} and \citet{lorch2012differential}.  If a person were to move from a home that is far from a specialty care provider to one that is near to a specialty care provider, whether the person would go the specialty care provider might depend on additional aspects of the move, such as the location and availability of public transportation at the new home and the exact travel time to the specialty care provider from the new home \citep{joffe2011,baiocchi2014}.  Consequently, a person may not be able to be deterministically classified as a complier or not a complier -- the person may be a complier with respect to certain moves but not others.
\item {\it{Calendar time IV}}.  Variations in the use of one treatment over another over time could result from changes in guidelines, changes in formularies or reimbursement policies, changes in physician preference (e.g., due to marketing activities by drug makers), release of new effectiveness or safety information, or the arrival of new treatments to the market.  For example, \citet{mack2013} studied for colon cancer patients the use of oxiliplatin as part of a multi-agent chemotherapy regimen vs. 5-fluorouracil monotherapy.  After FDA approval, oxiliplatin disseminated to replace 5-fluoracil monotherapy as the standard of therapy. \citet{mack2013} use whether a patient was treated before or after FDA approval as a calendar time IV.   Suppose the FDA approved a new drug on January 1, 2014 and the IV is whether or not a patient was treated before or after January 1, 2014.  A patient might be a never taker if she was treated on January 15, 2014, before the new drug has had much time to disseminate but a complier if she was treated on March 1, 2014, when the new drug has had time to disseminate.
\item {\it{Genes as IVs}}.  Genetic variants are increasingly being used as IVs; this approach is called {\it{Mendelian randomization}} because it makes use of the random assignment of genetic variants conditional on parents' genes discovered by Mendel.  The IV is often a single nucleotide polymorphism (SNP).  Changes in the SNP itself may not affect the treatment.  Instead, genetic variation at another location on the same chromosome as the SNP might affect the treatment; say the SNP is located at $L$ and the actual location on the chromosome that affects the treatment is $L^{'}$.  Even if a person's exposure level would change as a result of a change in the genetic code at location $L^{'}$, whether a person is a complier with respect to a change in the SNP at $L$ depends on whether the change in the SNP leads to a change in the genetic code at $L$, which is randomly determined through the process of recombination \citep{joffe2011}.
\item {\it{Random judges as an IV}}.  A number of papers have used randomly assigned judges/examiners as an IV, for example \citet{kling2006} and \citet{nagin2013} used randomly assigned judges with different sentencing tendencies to assess the effect of incarceration on employment, earnings and re-offending, \citet{maestas2013} used randomly assigned disability insurance examiners with different propensities for allowing a person to be eligible for Social Security Disability Insurance (SSDI) to assess the effect of SSDI insurance on labor supply and \citet{doyle2007} used randomly assigned child protection investigators with different tendencies to place a child into foster care to assess the effect of foster care on juvenile delinquency, teen motherhood and employment.  Although legal formalism holds that a judge applies legal reasons to the facts of a case in a rational, mechanical and deliberative manner and would presumably reach the same decision each time the judge examines a case \citep{leiter2005}, legal realism holds that psychological, political and social factors play a role and that a judge may reach a different decision depending on these psychological, political and social factors present at the time of a decision \citep{frank1930}.  \citet{danziger2011} found that even the time of the day and how tired/hungry a judge is may play a big role in a judge's decision: the percentage of prisoners applying for parole that were granted parole by judges dropped from approximately 65\% right after a food break to nearly zero right before a food break.  In this setting, the IV, the assignment of a judge to consider a parole case, is not delivered in a uniform way, as the effect of a judge assignment also depends on the time of day that the judge considers the case.
\end{itemize}

In the above examples, the fact that the IV is not delivered in a uniform way raises several problems with applying the deterministic compliance class framework of \citet{angrist1996identification}.  First, the stable unit treatment value assumption assumed by \citet{angrist1996identification} says that there is only one version of the IV.  In the above examples, there are multiple versions of the IV because depending on the way the IV is delivered, the IV can have a different effect on the treatment.  Second, in some of the above examples, the IV itself does not have a causal effect on the treatment, but is instead a proxy for something that has a causal effect, e.g., the physician's last prescription is a proxy for the physician's current prescribing preference.  Third, deterministic monotonicity is likely violated in some of the above examples because in some of the ways the IV is delivered, some of the people will be defiers.

In this paper, we present a framework for IVs that are not delivered in a uniform way that allows for there to be different versions of the IV, for the IV to be a proxy and for there to be violations of deterministic monotonicity as long as a weaker condition called stochastic monotonicity holds.  Stochastic monotonicity says that if we classify the subjects into types that we shall define, there are more compliers than defiers in each subgroup of the same type of subjects.  Our key result is that in our framework, the Wald estimate still estimates a certain treatment effect that is a weighted average of the treatment effect among each of the subgroups, where the subgroups for whom the IV has a stronger effect on the treatment get more weight.  We call this weighted treatment effect the strength-of-IV weighted average treatment effect (SIVWATE).  Our framework generalizes the framework of \citet{angrist1996identification} in that when the IV is delivered in a uniform way, our framework becomes equivalent to that of \citet{angrist1996identification} and the SIVWATE is equal to the LATE.

Previous literature has considered relaxing some of the assumptions of the \citet{angrist1996identification} framework that are violated when the IV is not delivered in a uniform way.  \citet{hernan2006instruments} and \citet{chalak2013} have presented results for when the measured IV is a proxy for an underlying, possibly continuous IV; their frameworks assume that the underlying IV satisfies deterministic monotonicity whereas we allow for the underlying IV to violate deterministic monotonicity and only satisfy the weaker condition of stochastic monotonicity.  When the IV is not a proxy and does not have multiple versions, several authors have presented results that identify the LATE or variants of the LATE when deterministic monotonicity is violated. \citet{small2007b} and \citet{dinardo2011} have presented a stochastic monotonicity condition.  \citet{brookhart2007} presented a similar formula when heterogeneity of treatment effects is generated by an unmeasured variable. \citet{chaisemartin2014} has provided a relaxation of deterministic monotonicity under which if there is a subgroup of compliers that accounts for the same proportion as the defiers and has the same local average treatment effect, then the Wald estimate captures the LATE of the remaining part of the compliers.  \citet{angrist1996identification} presented a formula for the sensitivity of the Wald estimate to violations of deterministic monotonicity.  \citet{klein2010} introduced local violations of deterministic monotonicity and showed that the bias of the two stage least squares (Wald) estimate can be well approximated if such violations are small.  \citet{huber2012} considered a local monotonicity assumption which requires that there be only compliers or defiers conditional on each value of the outcome.  Finally, \citet{robins1994}, \citet{hernan2006instruments}, and \citet{Tan2010} developed an IV approach that relies on some parametric modeling assumptions, instead of deterministic monotonicity, to achieve identification of average treatment effects on the treated at different instrument levels.

The contributions of this paper to the literature are that we provide a unified framework, identification results and inference methods that address the three problems that arise when an IV is not delivered in a uniform way (there may be different versions of the IV, the IV may be a proxy and there may be violations of deterministic monotonicity). Previous literature to our knowledge has not considered the presence of these three problems simultaneously.

Section 2 provides a review of the deterministic compliance class framework and identification results.  Section 3 presents the stochastic compliance class framework, the stochastic monotonicity assumption and identification results.  Section 4 provides sensitivity analysis for violations of the stochastic monotonicity assumption.  Section 5 presents applications of using an IV to study the effectiveness of high level neonatal intensive care units.  Section 6 provides discussion.

\bf 2. Review of Deterministic Compliance Class Framework and Identification Results.\rm

\bf 2.1. Deterministic Compliance Class Framework \rm

We review the deterministic compliance class framework of \citet{angrist1996identification}, where we use the conditional on measured covariates version of these assumptions formulated by \citet{abadie2003}.  Let $Z$ be the IV, $D$ be the treatment, $Y$ be the outcome and ${\bf{X}}$ be measured covariates.  The IV $Z$ and the treatment $D$ are assumed to be binary. We refer to level $1$ of the IV as the encouraging level and $0$ as the not encouraging level, and we refer to level $1$ of $D$ as the treatment and $0$ as the control.  Let there be $N$ subjects. Let $D_i(z_1,\ldots ,z_N)$ be the potential treatment taken if the vector of IV assignments was set to $z_1,\ldots ,z_N$ and let $Y_i(d_1,\ldots ,d_N,z_1,\dots ,z_N)$ be the potential outcome for subject $i$ if the vector of IV assignments for subjects $1,\ldots ,N$ was set to $z_1,\ldots ,z_N$ and the vector of treatment assignment for subjects $1,\ldots ,N$ was set to $d_1,\ldots ,d_N$.  The observed treatment for subject $i$ is $D_i\equiv D_i(Z_1,\ldots ,Z_N)$ and the observed outcome for subject $i$ is $Y_i\equiv Y_i(Z_1,\ldots ,Z_N,D_1,\ldots ,D_N)$.  The assumptions for $Z$ to be a valid IV in the deterministic compliance class framework are the following:
\begin{enumerate}
\item[DCC-IVA1] {\it{Stable Unit Treatment Value Assumption (SUTVA) for both treatment and IV}}.  There is no interference between subjects and there are not different version of the IV or treatment so that $D_i(z_1,\ldots ,z_N)=D_i(z_1^{'},\ldots ,z_N^{'})$ if $z_i=z_i^{'}$ and $Y_i(z_1,\ldots ,z_N,d_1,\ldots ,d_N)=Y_i(z_1^{'},\ldots ,z_N^{'},d_1^{'},\ldots ,d_N^{'})$ if $z_i=z_i^{'}$ and $d_i=d_i^{'}$.  Under SUTVA, we can write the potential treatments for subject $i$ as $D_i(z=1)$ and $D_i(z=0)$ where these are the treatments subject $i$ would have if $Z_i$ was set to $1$ and $0$ respectively, and we can write the potential outcomes as $Y_i(d=0,z=0), Y_i(d=0,z=1), Y_i(d=1,z=0), Y_i(d=1,z=1)$ where these are the potential outcomes subject $i$ would have if $D_i$ and $Z_i$ were set to the values $d$ and $z$ respectively.
    \item[DCC-IVA2] {\it{IV affects the treatment}}.  $P(D(1)=1|{\bf{X}})>P(D(0)=1|{\bf{X}})$.
    \item[DCC-IVA3] {\it{Independence of the instrument}}.  Conditional on ${\bf{X}}$, the vector $\{Y(d=0,z=0),Y(d=0,z=1),Y(d=1,z=0),Y(d=1,z=1),D(z=0),D(z=1)\}$ for a randomly drawn subject from the population is independent of $Z$.
        \item[DCC-IVA4] {\it{Exclusion restriction}}.  The IV has no direct effect on the outcome, only affecting the outcome through the treatment, so that $Y(d=0,z=0)=Y(d=0,z=1)$ and $Y(d=1,z=0)=Y(d=1,z=1)$.  Consequently, we write $Y(d=0)$ for $Y(d=0,z=0)=Y(d=0,z=1)$ and $Y(d=1)$ for $Y(d=1,z=0)=Y(d=1,z=1)$.
            \item[DCC-IVA5] {\it{Deterministic Monotonicity}}.  $D(z=1)\geq D(z=0)$ for all subjects.
\end{enumerate}

A subject's compliance class $C$ is $C=nt$ (never takers) if $D(z=0)=0,D(z=1)=0$, $C=at$ (always takers) if $D(z=0)=1,D(z=1)=1$, $C=co$ (compliers) if $D(z=0)=0,D(z=0)=1$ and $C=de$ (defiers) if $D(z=0)=1,D(z=0)=0$.  The deterministic monotonicity assumption (DCC-IVA5) says that there are no defiers.

\bf 2.2 Identification Results Under the Deterministic Compliance Class Framework \rm

\citet{imbens1994} and \citet{angrist1996identification} showed that under (DCC-IVA1)-(DCC-IVA5), average treatment effects for compliers are identified.  Specifically,
the conditional version of Proposition 1 in \citet{angrist1996identification} gives
\begin{equation}
E[Y(1)-Y(0)|C=co,{\bf{X}}]=\frac{E[Y|Z=1,{\bf{X}}]-E[Y|Z=0,{\bf{X}}]}{P(D=1|Z=1,{\bf{X}})-P(D=0|Z=1,{\bf{X}})}. \label{late.covariates}
\end{equation}
The proportion of compliers are identified by
 (\citet{abadie2003}, Lemma 2.1)
\begin{equation}
P(C=co|{\bf{X}})=P(D=1|Z=1,{\bf{X}})-P(D=1|Z=0,{\bf{X}}) \label{proportion.compliers.covariates}
\end{equation}
The overall average treatment effect for compliers, the LATE, is identified by
\citep{Tan2006}
\begin{equation}
E[Y(1)-Y(0)|C=co]=\frac{E[E(Y|Z=1,{\bf{X}})-E(Y|Z=0,{\bf{X}})]}{E[P(D=1|Z=1,{\bf{X}})-P(D=1|Z=0,{\bf{X}})]}. \label{overall.late}
\end{equation}

We now discuss estimation of the LATE from a sample of $N$ subjects.  When there are no measured covariates that need to be conditioned on for (DCC-IVA1)-(DCC-IVA5) to hold (for example, the IV is a randomized encouragement), then the LATE is
\begin{equation}
E[Y(1)-Y(0)|C=co]=\frac{E(Y|Z=1)-E(Y|Z=0)}{P(D=1|Z=1)-P(D=1|Z=0)} \label{wald.estimand}
\end{equation}
and the LATE is typically estimated by the sample analogue of (\ref{wald.estimand}),
\begin{equation}
\frac{\hat{E}(Y|Z=1)-\hat{E}(Y|Z=0)}{\hat{P}(D=1|Z=1)-\hat{P}(D=1|Z=0)}\label{wald.estimator}.
\end{equation}
(\ref{wald.estimator}) is called the Wald (two-stage least squares) estimator.  When there are measured covariates that need to be conditioned on for (DCC-IVA1)-(DCC-IVA5) to hold, the LATE (\ref{overall.late}) can be estimated by modeling $E(Y|Z=1,{\bf{X}})$, $E(Y|Z=0,{\bf{X}})$, $P(D=1|Z=1,{\bf{X}})$ and $P(D=1|Z=0,{\bf{X}})$ and then estimating the LATE by
\citep{frolich2007}
\begin{equation}
\frac{\sum_{i=1}^N \hat{E}(Y|Z=1,{\bf{X}}={\bf{X}}_i)-\hat{E}(Y|Z=0,{\bf{X}}={\bf{X}}_i)}{\sum_{i=1}^N \hat{P}(D|Z=1,{\bf{X}}={\bf{X}}_i)-\hat{P}(D|Z=0,{\bf{X}}={\bf{X}}_i)} \label{late.covariate.est}
\end{equation}
\citet{abadie2003}, \citet{Tan2006}, \citet{uysal2011} and \citet{ogburn2014} have developed other methods for estimating the LATE.

\bf 3. Stochastic Compliance Class Framework and Identification Results \rm

\bf 3.1 Stochastic Compliance Class Framework \rm

In the stochastic compliance class framework, we do not assume that a subject's compliance behavior (treatment taken as the level of the IV varies) is deterministic or even that it is well defined.  We allow for the IV to be a proxy, i.e., the IV does not have to causally affect the treatment itself and is  just associated with an underlying cause of the treatment.  We regard the IV $Z_i$ as fixed for subject $i$ and consider potential outcomes that vary $D$; this differs from the deterministic compliance class framework where potential outcomes vary with both $Z$ and $D$.  Let $Y_i(d_1,\ldots ,d_N)$ be the potential outcome for subject $i$ if the vector of treatment assignment for subjects $1,\ldots ,N$ was set to $d_1,\ldots ,d_N$, where the IVs $Z_i$ and the measured covariates ${\bf{X}}_i$ are pre-treatment covariates that remain the same for subject $i$ under all the $2^N$ potential outcomes for subject $i$.  Let ${\bf{U}}$ be a sufficient set of unmeasured common causes of $D$ and $Y$ such that conditional on ${\bf{U}}$, ${\bf{X}}$ and $Z$, the effect of $D$ on $Y$ is unconfounded, meaning
\begin{equation}
Y(d_1,\ldots ,d_N) \indep D|Z,{\bf{U}},{\bf{X}} \label{u.condition.before.sutva}
\end{equation}
for all the $2^N$ potential outcomes \citep{vanderweele2013b}.  The assumptions for a valid IV in the stochastic compliance class framework are the following:
\begin{enumerate}
\item[SCC-IVA1] {\it{Stable Unit Treatment Value Assumption (SUTVA) for treatment}}.  There is no interference between subjects and there are not different versions of treatment so that $Y_i(d_1,
    \ldots ,d_N)=Y_i(d_1
    ^{'},\ldots ,d_N^{'})$ if $d_i=d_i^{'}$.  Under SUTVA, we can write the potential outcomes for subject $i$ as $Y_i(d=1)$ and $Y_i(d=0)$ where these are the outcomes subject $i$ would have if $D$ was set to $1$ and $0$ respectively.  Also, under SUTVA, we can write (\ref{u.condition.before.sutva}) as
\begin{equation}
Y(d=0),Y(d=1)\indep D|Z,{\bf{U}},{\bf{X}}. \label{u.condition.after.sutva}
\end{equation}
\item[SCC-IVA2] {\it{IV is positively associated with treatment for at least some ${\bf{X}}$}}.  $P(D=1|Z=1,{\bf{X}})> P(D=1|Z=0,{\bf{X}})$ for at least one ${\bf{X}}$.
\item[SCC-IVA3] {\it{IV is conditionally independent of potential outcomes and unmeasured common causes ${\bf{U}}$ of $D$ and $Y$ given the measured covariates ${\bf{X}}$}}:  $(Y(d=0),Y(d=1),{\bf{U}})\indep Z|{\bf{X}}$.  Figure 1 is a directed acyclic graph depicting the SCC-IVA3 assumption; the conditional independence of the IV from potential outcomes and ${\bf{U}}$ given ${\bf{X}}$ comes from the lack of an arrow from ${\bf{U}}$ to $Z$ and the lack of an arrow from $Z$ to $Y$.
\item[SCC-IVA4] {\it{Stochastic Monotonicity}}.  $P(D=1|Z=1,{\bf{X}}={\bf{x}},{\bf{U}}={\bf{u}})\geq P(D=1|Z=0,{\bf{X}}={\bf{x}},{\bf{U}}={\bf{u}})$ for all ${\bf{x}}$, ${\bf{u}}$.
This means that the probability of having the treatment is at least as high for subjects with the encouraging level of the IV compared to the not encouraging level of the IV within all strata of ${\bf{X}},{\bf{U}}$.
\end{enumerate}
Note that there may be more than one sufficient set of unmeasured common causes of $D$ and $Y$ satisfying (\ref{u.condition.before.sutva}) \citep{vanderweele2013b}, but we say that $Z$ is a valid IV in the stochastic compliance class framework if it satisfies (SCC-IVA1)-(SCC-IVA4) for any sufficient set of unmeasured common causes ${\bf{U}}$ of $D$ and $Y$.  If $Z$ is a valid IV in the stochastic compliance class framework, then the causal effects we shall define in Section 3.2 are the same for all sufficient sets of unmeasured common causes ${\bf{U}}$ for which (SCC-IVA1)-(SCC-IVA4) are satisfied.

We now compare the assumptions for a valid IV in the stochastic compliance class framework to those in the deterministic compliance class framework. We note that when there are deterministic compliance classes, then the compliance classes (i.e., $\{D(z=0),D(z=1)\}$) are a sufficient set of unmeasured common causes of ${\bf{U}}$ of $D$ and $Y$ since $\{Y(d=0),Y(d=1)\} \indep D | \{D(z=0),D(z=1),{\bf{X}},Z\}$. In fact, $\{D(z=0),D(z=1)\}$ and $Z$ jointly determine $D$ in the deterministic compliance class framework.   Assumption SCC-IVA1 is similar to assumption DCC-IVA1 and when there are deterministic compliance classes, DCC-IVA1 implies SCC-IVA1.  Assumption SCC-IVA2 is similar to assumption DCC-IVA2 but is weaker in that it does not require the IV to have a causal effect on the treatment, only to be associated with the treatment.  When there are deterministic compliance classes, assumptions DCC-IVA3 and DCC-IVA4 together imply SCC-IVA3 because DCC-IVA3 and DCC-IVA4 together imply
\begin{equation}
\{Y(d=0),Y(d=1),D(z=0),D(z=1) \}\indep Z|{\bf{X}}, \label{dcciva3.dcciva4}
\end{equation}
and for the deterministic compliance class framework, we can take ${\bf{U}}=\{ D(z=0),D(z=1)\}$ so that (\ref{dcciva3.dcciva4}) is equivalent to SCC-IVA3.
The stochastic monotonicity assumption SCC-IVA4 is weaker than the deterministic monotonicity assumption DCC-IVA5 for the following reason.  Suppose there are deterministic compliance classes and DCC-IVA5 holds.   Then, SCC-IVA4 holds because for ${\bf{U}}=\{ D(z=0),D(z=1)\}$, we have $P(D=1|Z=1,{\bf{X}},{\bf{U}})\geq P(D=1|Z=1,{\bf{X}},{\bf{U}})$ by DCC-IVA5.  The stochastic monotonicity assumption SCC-IVA4 is strictly weaker than the deterministic monotonicity assumption DCC-IVA5 because the stochastic monotonicity assumption allows there to be defiers as long as within each strata of ${\bf{X}},{\bf{U}}$, there are at least as many compliers as defiers.  For example, in Table 1, if the previous patient (Alice, Bob or Carol) seen by Dr. X and Dr. Y is randomly determined with uniform probability, then for patients like Dave, there is a possibility of being a defier but it is more likely that Dave is a complier than a defier so that stochastic monotonicity is satisfied.

In summary, whenever there are deterministic compliance classes and $Z$ is a valid IV in the deterministic compliance class framework satisfying (DCC-IVA1)-(DCC-IVA5), then the stochastic compliance class framework also holds and $Z$ is also a valid IV under the stochastic compliance class framework satisfying (SCC-IVA1)-(SCC-IVA4).  However, the stochastic compliance class framework  and IV assumptions are more broadly applicable: the stochastic compliance class IV assumptions (SCC-IVA1)-(SCC-IVA4) are weaker than the deterministic compliance class framework IV assumptions (DCC-IVA1)-(DCC-IVA5) and also the stochastic compliance class framework can be used in settings where the deterministic compliance class framework does not apply such as when the compliance class is not well defined or the IV is a proxy.

\begin{figure}
\center{\scalebox{.5}{
\includegraphics{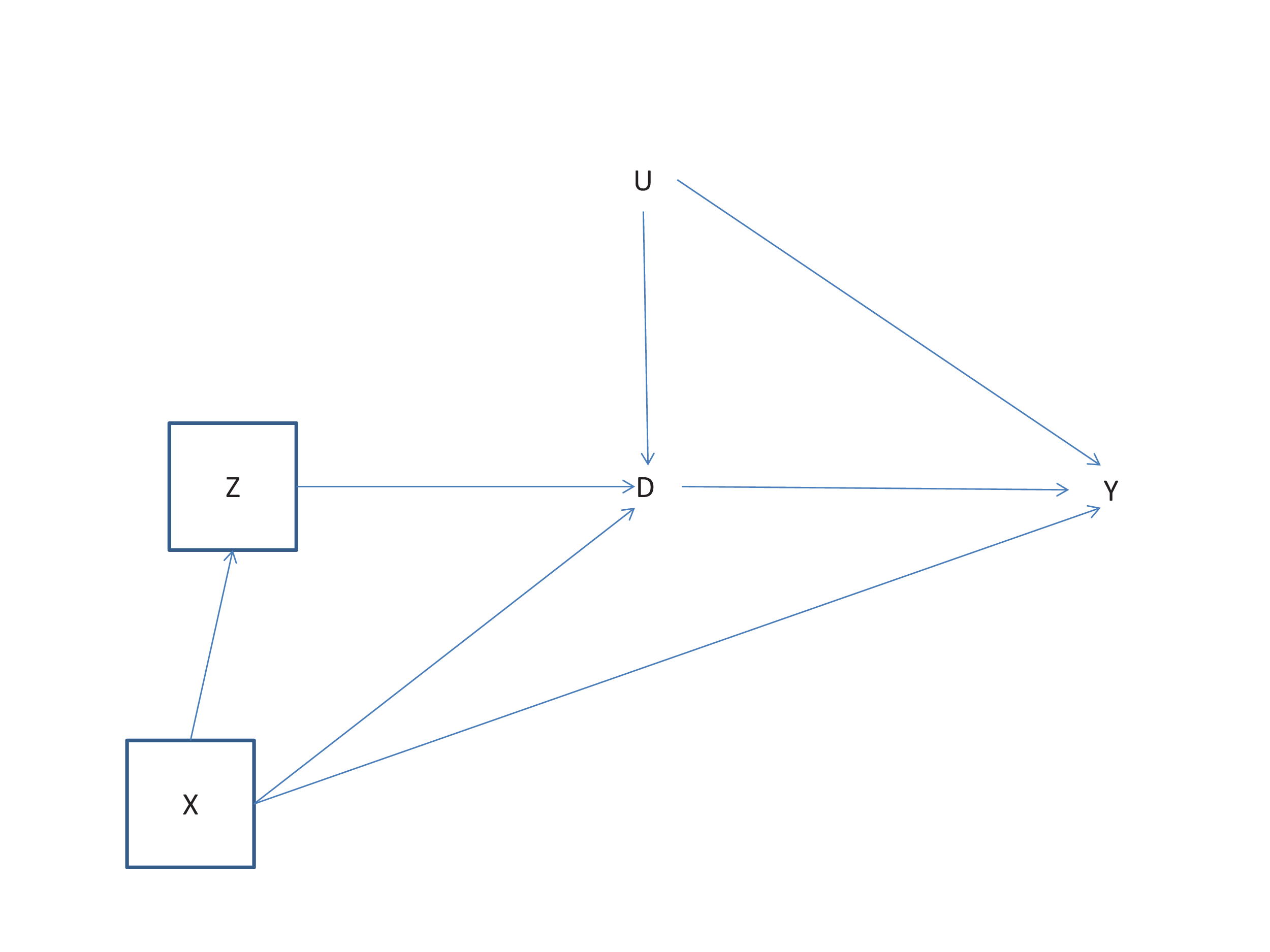}}}
\caption{Directed acyclic graph depicting the SCC-IVA3 assumption.  $Z$ is the IV, $D$ is the treatment, $Y$ is the outcome, $X$ are measured covariates and $U$ are unmeasured common causes of $D$ and $Y$.  The boxes around $Z$ and $X$ indicated that they are being conditioned on.  }
\label{dag}
\end{figure}


\bf 3.2 Identification Results Under the Stochastic Compliance Class Framework \rm

Let ${\mathcal{Q}}$ denote the weighted distribution from the population with the weight proportional to
$$
w({\bf{x}} , {\bf{u}}) = P(D=1|Z=1,{\bf{X}}={\bf{x}},{\bf{U}}={\bf{u}})-P(D=1|Z=0,{\bf{X}}={\bf{x}},{\bf{U}}={\bf{u}})
$$
for a unit with $\{ {\bf{X}}={\bf{x}}, {\bf{U}}={\bf{u}}\}$. Then ${\mathcal{Q}}$ samples more heavily from strata of ${\bf{U}}$ in which the IV is more associated with treatment.  Since ${\mathcal{Q}}$ weights subjects by how strongly the IV is associated with the treatment in that subject's subgroup (where the subgroups are defined by ${\bf{U}},{\bf{X}}$), we call the average treatment effect under ${\mathcal{Q}}$, the {\it{{Strength-of-IV Weighted Average Treatment Effect (SIVWATE)} }}:
\begin{equation}
E_{{\mathcal{Q}}}[Y(1)-Y(0)]
=\frac{\int E[Y(1)-Y(0)|{\bf{U}},{\bf{X}}] w({\bf{x}}, {\bf{u}}) dF({\bf{x}},{\bf{u}}) }{\int w({\bf{x}}, {\bf{u}}) dF({\bf{x}},{\bf{u}})  },
\label{SIVWATE}
\end{equation}
where for compactness of notation, we write $Y(d=1)$ as $Y(1)$ and $Y(d=0)$ as $Y(0)$ in (\ref{SIVWATE}) and henceforth.  The SIVWATE equals the LATE if compliance classes are deterministic and (DCC-IVA1)-(DCC-IVA5) hold, since then for ${\bf{U}}=\{ D(z=0),D(z=1)\}$, we have
$w({\bf{x}} , {\bf{u}})=1$ for compliers and $= 0$ for always takers and never takers.

The following two propositions show that functions of potential outcomes under the weighted distribution ${\mathcal{Q}}$, in particular the SIVWATE, are identified by a valid IV satisfying (SCC-IVA1)-(SCC-IVA4).
The proofs are provided in the appendix.  Proposition 1 involves identification in terms of the regression function of $(Y,D)$ given $(Z,\bf X)$, whereas
Proposition 2 involves identification in terms of the instrument propensity score $P(Z=1 | \bf X)$,
corresponding to, respectively, the regression and weighting methods in Tan (2006) under deterministic monotonicity.

\begin{proposition} Assume (SCC-IVA1)-(SCCIV-A4) holds.
For any measurable function $g$ with $E|g(Y(1))|<\infty$,
\begin{eqnarray}
E_{{\mathcal{Q}}}[g(Y(1))] & = & \frac{ E[ E(Dg(Y)|Z=1,{\bf{X}})-E(Dg(Y)|Z=0,{\bf{X}})]}{E[ P(D=1|Z=1,{\bf{X}})-P(D=1|Z=0,{\bf{X}}) ]}, \label{prop1.eqn1}
\end{eqnarray}
and for any measurable function $g$ with $E|g(Y(0))|<\infty$,
\begin{eqnarray}
E_{{\mathcal{Q}}}[g(Y(0))] & = & - \frac{ E[ E((1-D)g(Y)|Z=1,{\bf{X}}) -E((1-D)g(Y)|Z=0,{\bf{X}})]}{E[ P(D=1|Z=1,{\bf{X}})-P(D=1|Z=0,{\bf{X}}) ]}. \label{prop1.eqn2}
\end{eqnarray}
As a result,
\begin{eqnarray}
E_{{\mathcal{Q}}}[g(Y(1))- g(Y(0))] & = & \frac{ E[ E(g(Y)|Z=1,{\bf{X}})-E(g(Y)|Z=0,{\bf{X}})]}{E[ P(D=1|Z=1,{\bf{X}})-P(D=1|Z=0,{\bf{X}}) ]}. \label{prop1.eqn3}
\end{eqnarray}
\label{main.proposition}
\end{proposition}

\begin{corollary}
Assume (SCC-IVA1)-(SCCIV-A4) holds. For a covariate $V=V({\bf{X}})$ that is a function of ${\bf{X}}$, the conditional expectation of $g(Y(1))-g(Y(0))$ given $V$ under the weighted distribution $\mathcal Q$ is
\begin{eqnarray}
E_{{\mathcal{Q}}}[g(Y(1))- g(Y(0)) | V ] & = & \frac{ E[ E(g(Y)|Z=1,{\bf{X}})-E(g(Y)|Z=0,{\bf{X}}) |V ]}{E[ P(D=1|Z=1,{\bf{X}})-P(D=1|Z=0,{\bf{X}}) |V ]} \label{corollary1.eqn1}.
\end{eqnarray}
Moreover, we have
\begin{equation}
E_{{\mathcal{Q}}}[V] = \frac{E[ V (P(D=1|Z=1,{\bf{X}})-P(D=1|Z=0,{\bf{X}})) ]}{E [P(D=1|Z=1,{\bf{X}})-P(D=1|Z=0,{\bf{X}})]} \label{weighted.mean.x}
\end{equation}
\label{corollary.SIVWATE}
\end{corollary}

\begin{proposition}
Assume that (SCC-IVA1)-(SCCIV-A4) holds and that $0< P(Z=1|\bf X = \bf x) <1$ for almost all $\bf x$.  Define
\begin{eqnarray*}
\kappa & = &\frac{Z}{P(Z=1|{\bf{X}})} - \frac{1-Z}{P(Z=0|{\bf{X}})} \\
&=& \frac{Z-P(Z=1|{\bf{X}})}{P(Z=1|{\bf{X}}) P(Z=0|{\bf{X}})}.
\end{eqnarray*}
For any measurable function $g$ with $E|g(Y(1),{\bf{X}})|<\infty$,
\begin{eqnarray}
E_{{\mathcal{Q}}}[g(Y(1))] & = & \frac{E[\kappa \, D g(Y)]}{E(\kappa D)}, \label{y1.kappa.result}
\end{eqnarray}
and for any measurable function $g$ with $E|g(Y(0),{\bf{X}})|<\infty$,
\begin{eqnarray}
E_{{\mathcal{Q}}}[g(Y(0))] & = & -\frac{E[\kappa \, (1-D) g(Y)]}{E(\kappa D)}. \label{y0.kappa.result}
\end{eqnarray}
As a result,
\begin{eqnarray}
E_{{\mathcal{Q}}}[g(Y(1))- g(Y(0))] & = & \frac{E[\kappa \, g(Y)]}{E(\kappa D)} . \label{combined.kappa.result}
\end{eqnarray}
\label{kappa.proposition}
\end{proposition}

From Proposition \ref{main.proposition}, if (SCC-IVA1)-(SCC-IVA4) hold without conditioning any covariates ${\bf{X}}$, then the SIVWATE, $E_{{\mathcal{Q}}}[Y(1)-Y(0)]$ is
\begin{equation}
E_{{\mathcal{Q}}}[Y(1)-Y(0)]=\frac{E(Y|Z=1)-E(Y|Z=0)}{P(D=1|Z=1)-P(D=0|Z=0)}. \label{wald.estimand2}
\end{equation}
The right hand side of (\ref{wald.estimand2}) is identical to that of (\ref{wald.estimand}) under deterministic monotonicity, which is the probability limit of the Wald estimator (\ref{wald.estimator}).  Thus, Proposition \ref{corollary.SIVWATE} shows that if (SCC-IVA1)-(SCC-IVA4) hold without conditioning on any covariates ${\bf{X}}$ and we use the usual Wald (two stage least squares) estimator, then we obtain a consistent estimate of the SIVWATE.

When we do need to condition on covariates ${\bf{X}}$ for (SCC-IVA1)-(SCC-IVA4) to hold, the SIVWATE is (\ref{prop1.eqn3}) with $g(Y)=Y$; this expression is is identical to the expression (\ref{overall.late}) in \citet{Tan2006} for the LATE under deterministic monotonicity.  Similarly, (\ref{prop1.eqn1}) divided by (\ref{prop1.eqn2}) measures a multiplicative treatment effect under ${\mathcal{Q}}$ and is identical to the expression in \citet{ogburn2014} for the multiplicative LATE.  Equation (\ref{weighted.mean.x}) provides a formula for obtaining the mean of a covariate $V$ that is part of ${\bf{X}}$ under the weighted distribution ${\mathcal{Q}}$.
Finally, the weighting formulas (\ref{y1.kappa.result})-(\ref{combined.kappa.result}) in Proposition \ref{kappa.proposition} are analogues of previous deterministic compliance class framework results \citep{abadie2003,Tan2006} for the stochastic compliance class framework.


As we noted in Section 3.1, there may be more than one sufficient set of unmeasured common causes ${\bf{U}}$ for which $Z$ is a valid IV satisfying (SCC-IVA1)-(SCC-IVA4).  However, Propositions 1 and 2 apply to all such ${\bf{U}}$ and show that the SIVWATE is identified in terms of the observable $Z,D,Y,{\bf{X}}$ so that the SIVWATE is the same regardless of which ${\bf{U}}$ is considered.

When treatment effects are heterogeneous, \citet{imbens1994} and \citet{angrist1996identification} showed that the probability limit of the Wald estimator (\ref{wald.estimand}) has a disturbing sign reversal property: it is possible for the treatment effect to be positive for every subject but for the Wald estimator to converge in probability to a negative number.  However, under deterministic monotonicity, if the sign of the treatment effects (+, 0 or -) is the same for every subject in the population, then the sign of the treatment effects is identified because the sign of the probability limit of the Wald estimator (\ref{wald.estimand}) is equal to the average treatment effect for compliers.  Proposition 2 shows that this no sign reversal property also holds under stochastic monotonicity (assuming that (SCC-IVA1)-(SCC-IVA3) also hold): if the sign of the treatment effects (+, 0 or -) is the same for every subject in the population, the sign of the treatment effects is identified because the identified SIVWATE is a weighted average of treatment effects.

\bf 3.3. Bounds on the global average treatment effect \rm

Under the stochastic monotonicity assumption, a valid IV identifies a weighted average of treatment effects, the SIVWATE (Proposition \ref{main.proposition}).  The IV does not identify the unweighted, global average treatment effect, $E[Y(1)-Y(0)]$, but if a researcher is able to put bounds on how much the average treatment effect varies as ${\bf{X}}$ and ${\bf{U}}$ vary, then knowing the SIVWATE will provide bounds on the global average treatment effect.

\begin{proposition}
Suppose $\max_{\bf{u}} E[Y(1)-Y(0)|{\bf{U}}={\bf{u}},{\bf{X}}={\bf{x}}]-\min_{\bf{u}} E[Y(1)-Y(0)|{\bf{U}}={\bf{u}},{\bf{X}}={\bf{x}}]=r$ and that (SCC-IVA1)-(SCC-IVA4) hold.  Then the following are sharp bounds on $E[Y(1)-Y(0)|{\bf{X}}]$:
\begin{gather}
E_{{\mathcal{Q}}}[Y(1)-Y(0)|{\bf{X}}]-r(1-[P(D=1|Z=1,{\bf{X}})-P(D=1|Z=0,{\bf{X}})])\leq E[Y(1)-Y(0)|{\bf{X}}] \leq \nonumber \\
E_{{\mathcal{Q}}}[Y(1)-Y(0)|{\bf{X}}]+r(1-[P(D=1|Z=1,{\bf{X}})-P(D=1|Z=0,{\bf{X}})]) \label{bound}
\end{gather}
\end{proposition}

These are the same bounds as hold in the deterministic compliance class framework under assumptions (DCC-IVA1)-(DCC-IVA5) and the assumption that the maximum difference between the average treatment effect for compliers and the average treatment effect for always takers and never takers is $r$, with $E[Y(1)-Y(0)|C=co,{\bf{X}}]$ in place of $E_{{\mathcal{Q}}}[Y(1)-Y(0)|{\bf{X}}]$, see \citet{baiocchi2014}, Section 5.1.  The inequality (\ref{bound}) can be integrated over the distribution of ${\bf{X}}$ to obtain bounds on $E[Y(1)-Y(0)]$ (see Table 4 for an example)

\bf 4. Sensitivity Analysis for Violations of Stochastic Monotonicity\rm

Regardless of whether stochastic monotonicity holds, as long as (SCC-IVA1)-(SCC-IVA3) hold, the quantity on the right hand side of (\ref{prop1.eqn3}) with $g(Y)=Y$ that we use to estimate the SIVWATE is equal to
\begin{gather}
\frac{E[E(Y|Z=1,{\bf{X}})-E(Y|Z=0,{\bf{X}})]}{E[P(D=1|Z=1,{\bf{X}})-P(D=1|Z=0,{\bf{X}})]} = \\
\frac{E[E(Y(1)-Y(0)|{\bf{X}},{\bf{U}})w({\bf{X}},{\bf{U}})]}{E[w({\bf{X}},{\bf{U}})]} \label{SIVWATE.without.monotonicity}
\end{gather}
since the proof of Proposition \ref{main.proposition} did not make use of stochastic monotonicity.
When stochastic monotonicity is violated, (\ref{SIVWATE.without.monotonicity}) is not a weighted average of treatment effects because some of the ``weights'' $w({\bf{X}},{\bf{U}})$ are negative.  In this case, we might be interested in the strength of IV weighted average of treatment effects among subjects for whom the weights $w({\bf{X}},{\bf{U}})$ are positive, which we call the positive strength of IV weighted average treatment effect (PSIVWATE).  The PSIVWATE is equal to the following, where we let ${\mathcal{A}}=\{ ({\bf{X}},{\bf{U}}): w({\bf{X}},{\bf{U}})\geq 0\}$,
\begin{gather}
PSIVWATE = E_{{\mathcal{Q}}}[Y(1)-Y(0)|({\bf{X}},{\bf{U}})\in {\mathcal{A}}]= \nonumber \\
E\left[E[Y(1)-Y(0)|{\bf{X}},{\bf{U}}]\frac{w({\bf{X}},{\bf{U}})1\{ ({\bf{X}},{\bf{U}})\in {\mathcal{A}}\}}{E[w({\bf{X}},{\bf{U}})1\{ ({\bf{X}},{\bf{U}})\in {\mathcal{A}}\}]}\right],
\label{pSIVWATE}
\end{gather}
where $1\{ \cdot \}$ denotes the indicator function.
When stochastic monotonicity holds, the PSIVWATE is the SIVWATE and equals the right hand side of (\ref{prop1.eqn3}) with $g(Y)=Y$.
When stochastic monotonicity does not hold, then the following theorem gives the asymptotic bias from using the sample analogue of the right hand side of (\ref{prop1.eqn3}) with $g(Y)=Y$ to estimate the PSIVWATE, where we define the negative strength of IV weighted average treatment effect (NSIVWATE) as the weighted average treatment effect among subjects for whom the $w({\bf{X}},{\bf{U}})$ are negative and the subjects are weighted by the absolute value of $w({\bf{X}},{\bf{U}})$,
\begin{gather}
NSIVWATE = E_{{\mathcal{Q}}}[Y(1)-Y(0)|({\bf{X}},{\bf{U}})\in {\mathcal{A}}^C]= \nonumber \\ E\left[E[Y(1)-Y(0)|{\bf{X}},{\bf{U}}]\frac{w({\bf{X}},{\bf{U}})1\{ ({\bf{X}},{\bf{U}})\in {\mathcal{A}}^C\}}{E[w({\bf{X}},{\bf{U}})1\{ ({\bf{X}},{\bf{U}})\in {\mathcal{A}}^C\}]}\right]
\label{nSIVWATE}
\end{gather}

\begin{proposition}
When (SCC-IVA1)-(SCC-IVA3) holds but the stochastic monotonicity condition (SCC-IVA4) may not hold,
\begin{gather}
\frac{E[E(Y|Z=1,{\bf{X}})-E(Y|Z=0,{\bf{X}})]}{E[P(D=1|Z=1,{\bf{X}})-P(D=1|Z=0,{\bf{X}})]}-PSIVWATE= \nonumber \\
-\lambda (NSIVWATE-PSIVWATE), \label{stochastic.monotonicity.violation.bias}
\end{gather}
where
\[
\lambda = -\frac{E[w({\bf{X}},{\bf{U}})1\{({\bf{X}},{\bf{U}})\in {\mathcal{A}}^C\}]}{E[w({\bf{X}},{\bf{U}})]}.
\]
\label{proposition.stoch.mon.violation}
\end{proposition}

Proposition \ref{proposition.stoch.mon.violation} generalizes the formula in \citet{angrist1996identification} for the bias from using the Wald estimate to estimate the LATE when there are defiers (Proposition 3 in \citet{angrist1996identification}); when the DCC framework holds, then (\ref{stochastic.monotonicity.violation.bias}) is equal to the bias formula in \citet{angrist1996identification}.  The bias due to violations of stochastic monotonicity is composed of two factors.  The first factor $\lambda$ is related to the proportion of subjects for whom stochastic monotonicity is violated and is equal to zero under the stochastic monotonicity assumption.  The numerator of $\lambda$ relates to the proportion of subjects for whom stochastic monotonicity is violated and the magnitude by which stochastic monotonicity is violated for these subjects, the smaller this proportion and magnitude, the smaller the numerator will be.  The denominator of $\lambda$ is equal to the overall strength of the association between the IV and the treatment, $E[w({\bf{X}},{\bf{U}})]=E[P(D=1|Z=1,{\bf{X}})-P(D=1|Z=0,{\bf{X}})]$ (see (\ref{proof.equation.minus2})).  The stronger the IV is, the less sensitive the IV estimate is to violations of stochastic monotonicity.  The second factor in the bias formula, $PSIVWATE-NSIVWATE$, is related to the difference in treatment effects between those subjects for whom treatment is positively associated with the IV and those subjects for whom treatment is negatively associated with the IV.  The less difference there is between treatment effects for these two types of subjects, the less bias there is from violations of the stochastic monotonicity assumption.

\bf 5. Application to study the effectiveness of high-level neonatal intensive care units\rm

\citet{lorch2012differential} studied the effect for premature babies of being delivered in a high level neonatal intensive care unit (high-level NICU) vs. a low-level NICU where a high-level NICU is defined as a NICU that has the capacity for sustained mechanical assisted ventilation and delivers at least 50 premature babies per year.  We consider the outcome of mortality.  We consider here data from Pennsylvania from 1995-2005 (192,078 premature babies).  The data was collected from birth and death certificates and the UB-92 form that hospitals use for billing purposes.  A baby's health status before delivery is an important confounder as mothers are more likely to go to a high-level NICU if a baby is considered to be at high risk for complications or death.  The data contains some measures of the baby's health such as gestational age, but the data is also missing several important measures such as fetal heart tracing results, the severity of maternal problems during pregnancy (e.g., the data contains an indicator for whether a mother had pregnancy-induced hypertension but no information on the severity) and the mother's adherence to prenatal guidelines.  Concern about these unmeasured confounders motivated \citet{baiocchi2010building} and \citet{lorch2012differential} to use an IV approach, using the mother's excess travel time from the nearest high-level NICU compared to the nearest low-level NICU as an IV.

Here we consider the IV to be $Z=1$ if the mother's excess travel time is less than or equal to 10 minutes and $Z=0$ if the mother's excess travel time is greater than 10 minutes.  The travel time is computed using Dijkstra's \citep{dijkstra1959} algorithm for computing the shortest path between two points as implemented in ArcView software.  This putative IV is likely to violate deterministic monotonicity for at least two reasons.  First, the travel time from the ArcView software may not accurately represent the travel time faced by the mother at the time when she needs to decide where to deliver because the travel time depends on the traffic at that time of day and also, if the mother uses public transportation, the travel time depends on public transportation routes.  Consequently, a mother who lives less than 10 minutes from the nearest high level NICU according to the ArcView software might actually have less travel time had she moved to a place that was further away according to the ArcView software but closer in terms of public transportation or traffic at the time she needs to go to the hospital.  Second, choice of hospital to deliver at is driven by a number of factors in addition to travel time and how high risk a patient is; these additional factors include where a patient's physician practices, the general view of the hospital by a specific community of patients and what the patient's family and friends believe about a hospital.  For example, a mother who would go to a high level NICU if she lived in a place where the excess travel time is 15 minutes might instead go to a low level NICU if she moved to a place where the excess travel time is 5 minutes if the patient's family and friends thought that that the low level NICU in the place where she moved to was a good place to have a baby.

Although deterministic monotonicity is not plausible, stochastic monotonicity (SCC-IVA4) is plausible.  The primary unmeasured common causes ${\bf{U}}$ of treatment (high-level NICU vs. low-level NICU) and outcome (mortality) are the baby's health prior to delivery.  Within each strata of health prior to delivery, it is plausible that differences between the ArcView travel time and actual travel time, and differences in factors like family and friends' beliefs about the hospitals in the area average out between the $Z=1$ and $Z=0$ patients so that within each strata of health prior to delivery, the patients with $Z=1$ (ArcView excess travel time $\leq$ 10 minutes) are more likely to go to a high level NICU than patients with $Z=0$ (ArcView excess travel time $>$ 10 minutes).  In addition to (SCC-IVA4) being plausible, the other assumptions needed for excess travel time to be a valid IV in the stochastic compliance class framework are plausible:
  \begin{itemize}
  \item SCC-IVA1.  The no interference part of SCC-IVA1 is plausible as baby $A$'s being treated at a high level NICU would usually not be affected by baby $B$'s outcome.  The no different versions of the treatment part of SCC-IVA1 is unlikely to hold exactly since different high level NICUs may differ in their level of care and different low level NICUs may differ
in their level of care, but it plausibly holds approximately; see \citet{vanderweele2013} for discussion about interpreting causal estimates
when there are different versions of treatment.
\item SCC-IVA2.  A mother typically obtains prenatal care from and would prefer to
deliver at a close by hospital so that a smaller excess travel time to the nearest high level NICU makes a mother more likely to deliver at a high level NICU \citep{phibbs1993choice}.
\item SCC-IVA3.  Women do not expect to have a premature delivery, and
thus conditional on measured socioeconomic variables such as mother's education, women do not choose where to
live based on distance to a high level NICU, making independence of excess travel time from unmeasured confounders plausible and most mothers have time to reach either the nearest high level or low level NICU before delivering so that the marginal travel time should not directly affect outcomes\citep{lorch2012differential}.
\end{itemize}

We estimated the SIVWATE by fitting (\ref{late.covariate.est}) using logistic regression where we use the measured covariates ${\bf{X}}$ listed in \citet{lorch2012differential} that include baby characteristics such as birthweight and gestational age; the mother's comorbidities such as gestational diabetes, diabetes mellitus, pregnancy induced hypertension and chronic hypertension; the mother's education, race and type of health insurance; and socioeconomic characteristics of the zip code the mother lives in such as the average income and percent below poverty.  Table 2 shows the overall SIVWATE estimates based on (\ref{main.proposition}) and the SIVWATE estimates based on (\ref{corollary1.eqn1}) for babies in three ranges of gestational age -- moderate to late preterm (33-37 weeks), very preterm (28-32 weeks) and extremely preterm ($\leq 27$ weeks).  The estimates are expressed in terms of the effect of high level NICUs vs. low level NICUs on mortality per 1000 births.  95\% confidence intervals were computed using the percentile bootstrap, with the resampling stratified on the three ranges of gestational ages.  The strength of IV weighted average treatment effect (SIVWATE) estimate is that being delivered in a high level NICU prevents 6 deaths per 1000 births with a 95\% confidence interval of preventing 2.5 to 8.7 deaths; our analysis indicates that high level NICUs
are effective.  The effect of high level NICUs on reducing mortality is estimated to be greater for more premature babies, with a particularly large effect for extremely premature babies ($\leq 27$ weeks).
\begin{table}[h!]
\begin{center}
\begin{tabular}{|c|c|c|}
\hline Group & Estimate & 95\% CI \\ \hline
All & -6.0 & (-8.7,-2.5) \\ \hline
Gestational Age, 33-37 wks & -1.8 & (-2.8,-0.7) \\
Gestational Age, 28-32 wks & -26.0 & (-38.5,-11.0) \\
Gestational Age, $\leq 27$ wks & -110.7 & (-164.8,-45.4) \\
\hline
\end{tabular}
\end{center}
\caption{SIVWATE estimates and confidence intervals for effect of delivering in high level NICUs vs. low level NICUs on mortality per 1000 premature births}
\label{SIVWATE.est.table}
\end{table}

Table 3 compares the distribution of characteristics in the strength of IV weighted population ${\mathcal{Q}}$ to the unweighted population using formula
(\ref{weighted.mean.x}), namely using the estimate
\[
\hat{E}_{{\mathcal{Q}}}[V] = \frac{\sum_{i=1}^N V_i[\hat{P}(D=1|Z=1,{\bf{X}}_i)-\hat{P}(D=1|Z=0,{\bf{X}}_i)]}{\sum_{i=1}^N \hat{P}(D=1|Z=1,{\bf{X}}_i)-\hat{P}(D=1|Z=0,{\bf{X}}_i)}.
\]
The strength of IV weighted population has more moderate to late premature babies (33-37 weeks) and less very (28-32 weeks) or extremely ($\leq 27$ weeks) premature babies.  Since the effect of high level NICUs appears to be greater among very and extremely premature babies (Table \ref{SIVWATE.est.table}), the SIVWATE for all babies may underestimate the global average treatment effect for all babies.  The strength of IV weighted population is similar to the full population in terms of mother's education, race and mother's comorbidities.
\begin{table}[h!]
\begin{center}
\begin{tabular}{|c|c|c|c|}
\hline Characteristic $X$ & Prevalence of $X$ & Prevalence of $X$ & Ratio of Prevalence \\
& in weighted & in unweighted & in ${\mathcal{Q}}$ \\
& population ${\mathcal{Q}}$ & population & to unweighted\\
\hline
Gestational age, 33-37 wks & 0.90 & 0.87 & 1.03 \\
Gestational age, 28-32 wks & 0.08 & 0.10 & 0.84 \\
Gestational age, $\leq 27$ wks & 0.02 & 0.03 & 0.58 \\
Birthweight $<$ 1500 g & 0.06 & 0.09 & 0.73 \\
Mother College Graduate & 0.26 & 0.26 & 0.99 \\
African American & 0.16 & 0.16 & 0.96 \\
Gestational Diabetes & 0.05 & 0.05 & 1.00 \\
Diabetes mellitus & 0.02 & 0.02 & 0.92 \\
Pregancy-induced hypertension & 0.10 & 0.10 & 0.95 \\
Chronic hypertension & 0.02 & 0.02 & 0.93 \\ \hline
\end{tabular}
\end{center}
\label{characteristic.table}
\caption{Characteristics of SIVWATE weighted population compared to unweighted population for NICU study}
\end{table}

We now consider the sensitivity of the SIVWATE estimates to violations of the stochastic monotonicity assumption using Proposition 4.  Proposition 4 provides information about how much bias there is in the estimates in Table 2 as estimates of the PSIVWATE, the strength of IV weighted average treatment effect for subjects whose chance of delivering at a high level NICU is positively affected by living near (as measured by the ArcView software) to a high level NICU.  We estimate the denominator of $\lambda$ (which equals $E[P(D=1|Z=1,{\bf{X}})-P(D=1|Z=0,{\bf{X}})]$ by $\sum_{i=1}^N \hat{P}(D=1|Z=1,{\bf{X}})-\hat{P}(D=1|Z=0,{\bf{X}})$, which equals 0.40 for the data set.  One way of interpreting this quantity is that if every baby has some probability of being a complier, defier, always taker or never taker -- say due to there being a number of possible new places the baby's mother could move which would change whether she was near vs. far from the high level NICU and the new place to move being chosen randomly accordingly to a probability distribution -- then for a randomly drawn baby and a random draw from that baby's compliance distribution of whether the baby is a complier, defier, always taker or never taker, the probability of being a complier minus the probability of being a defier is 40\%.  The numerator of $\lambda$ can be interpreted in the same way as the probability that a randomly drawn baby and a random draw from that baby's compliance distribution of whether the baby is a complier, defier, always taker or never taker results in a defier.
  Suppose this is 5\%, the same probability of defier considered in \citet{angrist1996identification}'s sensitivity analysis and a probability of defier that seems high in our setting.  Next, suppose that the difference between the PSIVWATE and the NSIVWATE is at most 24.2 deaths per 1000 births; this number is chosen because it is the difference between the estimated SIVWATEs for gestational age of 33-37 weeks and gestational age of 28-32 weeks in Table 2, two fairly different risk groups.  Then, by Proposition 4, the estimated overall PSIVWATE could range from $-6+(.05/.4)24.2=-3.0$ to $-6-(.05/.4)24.2=-9.0$.  Thus, the PSIVWATE estimate that delivering at a high-level NICU reduces mortality is robust to plausible amounts of departure from stochastic monotonicity.

We now consider putting bounds on the global average treatment effect using Proposition 3.    Suppose that for fixed ${\bf{X}}$ and varying ${\bf{U}}$, $E[Y(1)-Y(0)|{\bf{U}}={\bf{u}},{\bf{X}}={\bf{x}}]$ has a range from $(1/m)\times E_{{\mathcal{Q}}}[Y(1)-Y(0)|{\bf{X}}]$ to $m\times E_{{\mathcal{Q}}}[Y(1)-Y(0)|{\bf{X}}]$ so that $r=\left (m-\frac{1}{m}\right) \times E_{{\mathcal{Q}}}[Y(1)-Y(0)|{\bf{X}}]$ in Proposition 3.  Then, using Proposition 3, we estimate the bounds on $E[Y(1)-Y(0)|{\bf{X}}]$ to be
\begin{gather}
\mbox{Estimated Lower Bound} = \nonumber \\
\hat{E}_{{\mathcal{Q}}}[Y(1)-Y(0)|{\bf{X}}]\left( 1-m+\frac{1}{m}+\left(m-\frac{1}{m}\right)[P(D=1|Z=1,{\bf{X}})-P(D=1|Z=0,{\bf{X}})]\right)
\label{elb}
\end{gather}
to
\begin{gather}
\mbox{Estimated Upper Bound} = \nonumber \\
\hat{E}_{{\mathcal{Q}}}[Y(1)-Y(0)|{\bf{X}}]\left( 1+m-\frac{1}{m}-\left(m-\frac{1}{m}\right)[P(D=1|Z=1,{\bf{X}})-P(D=1|Z=0,{\bf{X}})]\right).
\label{eub}
\end{gather}
We have $E[Y(1)-Y(0)]=E[E[Y(1)-Y(0)|{\bf{X}}]]$, which we can estimate by $\frac{1}{N}\sum_{i=1}^N \hat{E}[Y(1)-Y(0)|{\bf{X}}]$.  Substituting the estimated lower and upper bounds (\ref{elb}) and (\ref{eub}) into this latter expression provides estimates of the bounds on $E[Y(1)-Y(0)]$.
\begin{table}[h!]
\label{bounds.table}
\begin{center}
\begin{tabular}{|c|c|c|}
\hline $m$ & Estimated Bounds & 95\% CI for bounds \\ \hline
1.1 & (-9.5, -8.6) & (-15.2, -3.7) \\
1.5 & (-11.0, -7.1) & (-17.7, -3.0) \\
2 & (-12.6, -5.5) & (-20.1, -2.3) \\
3 & (-15.3, -2.8) & (-24.8, -1.2) \\
5 & (-20.4, 2.2) & (-32.9, 3.6) \\ \hline
\end{tabular}
\end{center}
\caption{Bounds on global average treatment effect (mortality per 1000 births) when for fixed ${\bf{X}}$ and varying ${\bf{U}}$, $E[Y(1)-Y(0)|{\bf{U}}={\bf{u}},{\bf{X}}={\bf{x}}]$ has a range of $\left( m-\frac{1}{m}\right)\times E_{{\mathcal{Q}}}[Y(1)-Y(0)|{\bf{X}}]$.\label{bound.table}}
\end{table}
Table 4 shows the estimated bounds for $m=1.1,1.5,2,3,5$ along with 95\% confidence bounds formed by bootstrap resampling and using the Bonferroni method, which means taking the 2.5th percentile of the bootstrapped lower bound
and the 97.5th percentile of the upper bound \citep{horowitz2000,cheng2006}.  For moderate amounts of treatment effect heterogeneity, $m=1.1$ to $3$, the upper bound is below zero so that there is evidence that delivering all premature babies at high level NICUs compared to delivering all premature babies at low level NICUs would reduce mortality.


\bf 6. Discussion\rm

We have presented a framework, identification results and inference methods for using an IV to estimate a causal effect when an IV is not uniformly delivered and deteministic monotonicity may be violated.  Our framework generalizes the deterministic compliance class framework and identification results of \citet{angrist1996identification}.  \citet{angrist1996identification} showed that under the deterministic compliance class framework, when deterministic monotonicity holds, the standard IV estimators can be interpreted as estimating the LATE.  We have generalized this to that when stochastic monotonicity holds, the standard IV estimators can be interpreted as estimating the SIVWATE, the weighted average of treatment effects weighted by how strongly the IV is associated with a subject's treatment.  The SIVWATE is equal to the LATE when deterministic monotonicity holds.

\citet{angrist1996identification}'s work on the LATE has had a large influence on how researchers understand instrumental variable analyses and interpret what can be learned from instrumental variable analyses.  However, there has been controversy over whether the LATE is a useful estimand because (i) the LATE is an average treatment effect over a subpopulation, the compliers, that cannot be identified in the sense that there are no subjects whom we know belong to the subpopulation although there are some subjects whom we know do not belong to the subpopulation and (ii) the LATE is not usually the treatment effect of primary interest, instead the global average treatment effect over the whole population is often of greater interest \citep{freedman2006,deaton2010,pearl2011,swanson2014}.  Our generalization of the LATE, the SIVWATE, could be criticized along the same lines.  We support Imbens' discussion (Section 4.6 of \citet{imbens2014}) of why he feels estimating the LATE is useful and we feel similar reasoning also applies to the SIVWATE.  For studying a treatment, we would ideally like to have a randomized trial with perfect compliance.  IV analysis is only used when for practical or ethical reasons, we do not have this ideal study and instead have an observational study with unmeasured confounding (or a randomized trial with noncompliance); as Imbens says, ``IV analysis is an analysis in a second-best setting.''  Under deterministic monotonicity, the LATE tells us what we can learn directly about treatment effects from the data without making homogeneity assumptions about the treatment effect.  Under stochastic monotonicity, the SIVWATE tells us what we can learn directly about treatment effects from the data without making either homogeneity assumptions about the treatment effect or the assumption of deterministic monotonicity.  We can describe the weighted population that the SIVWATE (LATE) refers to in terms of the observed covariates by (\ref{weighted.mean.x}), as done in the Application in Section 5 (Table 3). The SIVWATE can be combined with assumptions about how heterogeneous treatment effects are to find bounds on the global average treatment effect (Section 3.3 and Table \ref{bound.table}).



\bf 7. Appendix\rm

\begin{proof}[Proof of Proposition 1]
For $z=0, 1$, we have
\begin{align*}
E[ g(Y) D | Z=z, {\bf{X}}] & = E[g(Y(1))D|Z=z,{\bf{X}}] \\
& = E[E[g(Y(1))D|{\bf{U}},Z=z,{\bf{X}},Y(1)]|Z=z,{\bf{X}}] \\
& = E[E[g(Y(1))P(D=1|{\bf{U}},Z=z,{\bf{X}},Y(1))|{\bf{U}},Z=z,{\bf{X}}]|Z=z,{\bf{X}}] \\
& = E[E[g(Y(1))P(D=1|{\bf{U}},Z=z,{\bf{X}})|{\bf{U}},Z=z,{\bf{X}}]|Z=z,{\bf{X}}] \\
& = E[g(Y(1))P(D=1|{\bf{U}},Z=z,{\bf{X}})|Z=z,{\bf{X}}] \\
& = E[g(Y(1))P(D=1|{\bf{U}},Z=z,{\bf{X}})|{\bf{X}}]
\end{align*}
where the fourth equality holds because $Y(1)\indep D|Z,{\bf{U}},{\bf{X}}$ from (\ref{u.condition.after.sutva}) and the last equality holds because $(Y(1),{\bf{U}})\indep Z|{\bf{X}}$ from SCC-IVA3.  Taking the difference of the preceding expressions with $z=1$ and $z=0$ gives
\begin{align}
E( g(Y) D | Z=1, {\bf{X}}) - E( g(Y) D | Z=0, {\bf{X}}) =
E[ g(Y(1)) w({\bf{X}},{\bf{U}}) | {\bf{X}} ].  \label{app-eqn1}
\end{align}
Taking $g(Y)$ to be a constant gives
\begin{align}
E( D | Z=1, {\bf{X}}) - E( D | Z=0, {\bf{X}}) =
E[  w({\bf{X}},{\bf{U}}) | {\bf{X}} ]. \label{app-eqn2}
\end{align}
Therefore, we have
\begin{align*}
\frac{ E( g(Y) D | Z=1, {\bf{X}}) - E( g(Y) D | Z=0, {\bf{X}}) }{ E( D | Z=1, {\bf{X}}) - E( D | Z=0, {\bf{X}}) } = \frac{E[ g(Y(1)) w({\bf{X}},{\bf{U}}) | {\bf{X}} ]} { E[ w({\bf{X}},{\bf{U}}) | {\bf{X}} ]}
\end{align*}
Taking the expectations of (\ref{app-eqn1}) and (\ref{app-eqn2}) over ${\bf{X}}$ leads to (\ref{prop1.eqn1}):
\begin{align*}
\frac{ E[E( g(Y) D | Z=1, {\bf{X}}) - E( g(Y) D | Z=0, {\bf{X}})] }{ E[ E( D | Z=1, {\bf{X}}) - E( D | Z=0, {\bf{X}}) ] } = \frac{E[ g(Y(1)) w({\bf{X}},{\bf{U}}) ]} { E[ w({\bf{X}},{\bf{U}})]},
\end{align*}
which is by definition $E_{\mathcal Q}[ g(Y(1))]$. Similarly, equation (\ref{prop1.eqn2}) can be proved.
\end{proof}

\begin{proof}[Proof of Corollary 1]
Taking the conditional expectations of (\ref{app-eqn1}) and (\ref{app-eqn2}) given $V$ leads to
\begin{align*}
\frac{ E[ E( g(Y) D | Z=1, {\bf{X}}) - E( g(Y) D | Z=0, {\bf{X}}) | V]}{ E[ E( D | Z=1, {\bf{X}}) - E( D | Z=0, {\bf{X}})|V] } = \frac{E[ g(Y(1)) w({\bf{X}},{\bf{U}}) | V ]} { E[ w({\bf{X}},{\bf{U}}) | V ]},
\end{align*}
which is $E_{\mathcal Q}( g(Y(1)) |V)$, the conditional expectation of $g(Y(1))$ given $V$ under the weighted distribution $\mathcal Q$. In fact,
for any function $h(V)$ such that $E_{\mathcal Q} |g(Y(1)) h(V) | < \infty$, we have
\begin{align*}
& E_{\mathcal Q} \left\{ \frac{E[ g(Y(1)) w({\bf{X}},{\bf{U}}) | V ]} { E[ w({\bf{X}},{\bf{U}}) | V ]} h(V) \right\}
= E_{\mathcal Q} \left\{ \frac{E[ g(Y(1)) w({\bf{X}},{\bf{U}})h(V) | V ]} { E[ w({\bf{X}},{\bf{U}}) | V ]} \right\} \\
& = E_{\mathcal{Q}} [E_{\mathcal{Q}}[g(Y(1))h(V)|V]]=E_{\mathcal{Q}}[g(Y(1))h(V)].
\end{align*}
Similarly, it can be shown that
\begin{align*}
E_{{\mathcal{Q}}}[g(Y(0)) |V] = - \frac{ E[ E((1-D)g(Y)|Z=1,{\bf{X}}) -E((1-D)g(Y)|Z=0,{\bf{X}}) |V]}{E[ P(D=1|Z=1,{\bf{X}})-P(D=1|Z=0,{\bf{X}}) |V ]}.
\end{align*}
and hence equation (\ref{corollary1.eqn1}) holds.
Moreover, equation (\ref{weighted.mean.x}) follows because $g(Y)$ can be generalized to a function of both $Y$ and ${\bf{X}}$ in the above proof.
\end{proof}

\begin{proof}[Proof of Proposition 2]
For $z=0, 1$, we have
\begin{align*}
& E\left[ \frac{1\{Z=z\}}{P(Z=z|{\bf{X}})} Dg(Y) |{\bf{X}} \right] = E\left[ E\left( \frac{1\{Z=z\}}{P(Z=z|{\bf{X}})} Dg(Y) | Z=z, {\bf{X}}\right) P(Z=z|{\bf{X}})\right]\\
& =E [ E(D g(Y) | Z=z, {\bf{X}}) ]
\end{align*}
by direct calculation.
Then equation (\ref{y1.kappa.result}) follows from the proof of Proposition 1. Similarly, equation (\ref{y0.kappa.result}) can be proved.
\end{proof}

\begin{proof}[Proof of Proposition 3]
We have
\begin{gather}
E[Y(1)-Y(0)|{\bf{X}}] = \nonumber \\
E[E[Y(1)-Y(0)|{\bf{X}},{\bf{U}}]|{\bf{X}}] = \nonumber \\
E[E[Y(1)-Y(0)|{\bf{X}},{\bf{U}}]w({\bf{X}},{\bf{U}})|{\bf{X}}] + \label{evy1y0.decomposition1} \\
E[E[Y(1)-Y(0)|{\bf{X}},{\bf{U}}]\{1-w({\bf{X}},{\bf{U}})\}|{\bf{X}}]. \label{evy1y0.decomposition2}
\end{gather}
We also have
\begin{gather}
E[w({\bf{X}},{\bf{U}})] = \nonumber \\
\int [P(D=1|Z=1,{\bf{X}},{\bf{U}})-P(D=1|Z=0,{\bf{X}},{\bf{U}})]dF({\bf{U}}|{\bf{X}}) = \nonumber \\
\int [P(D=1|Z=1,{\bf{X}},{\bf{U}})dF({\bf{U}}|{\bf{X}},Z=1)-P(D=1|Z=0,{\bf{X}},{\bf{U}})]dF({\bf{U}}|{\bf{X}},Z=0) = \nonumber \\
P(D=1|Z=1,{\bf{X}})-P(D=1|Z=0,{\bf{X}}).
\label{proof.equation.minus2}
\end{gather}
where the second equality follows from ${\bf{U}}\indep Z|{\bf{X}}$ from (SCC-IVA3).  Since $E_{{\mathcal{Q}}}[Y(1)-Y(0)|{\bf{X}}]-r\leq E[Y(1)-Y(0)|{\bf{X}},{\bf{U}}] \leq E_{{\mathcal{Q}}}[Y(1)-Y(0)|{\bf{X}}]+r$ for all ${\bf{U}}$ (by the assumptions of the theorem) and $E[\{ 1-w({\bf{X}},{\bf{U}})\}]= 1-[P(D=1|Z=1,{\bf{X}})-P(D=1|Z=0,{\bf{X}})]$ by (\ref{proof.equation.minus2}), we have that (\ref{evy1y0.decomposition2}) is lower bounded by $\{E_{{\mathcal{Q}}}[Y(1)-Y(0)|{\bf{X}}]-r\} \{1-[P(D=1|Z=1,{\bf{X}})-P(D=1|Z=0,{\bf{X}})]\}$ and upper bounded by $\{E_{{\mathcal{Q}}}[Y(1)-Y(0)|{\bf{X}}]+r\} \{1-[P(D=1|Z=1,{\bf{X}})-P(D=1|Z=0,{\bf{X}})]\}$.  Combining this with the fact that (\ref{evy1y0.decomposition1}) is equal to $E_{{\mathcal{Q}}}[Y(1)-Y(0)|{\bf{X}}]$ proves that the bounds in (\ref{bound}) hold.  These bounds are sharp because the lower bound in (\ref{bound}) is attained for the distribution of ${\bf{U}}|{\bf{X}}$ such that ${\bf{U}}$ takes on two values $0$ and $1$ with $P(D=1|Z=1,{\bf{X}},{\bf{U}}=1)-P(D=1|Z=1,{\bf{X}},{\bf{U}}=1)=1$, $P(D=1|Z=1,{\bf{X}},{\bf{U}}=0)-P(D=1|Z=1,{\bf{X}},{\bf{U}}=0)=0$,  $P({\bf{U}}=1|{\bf{X}})=P(D=1|Z=1,{\bf{X}})-P(D=1|Z=0,{\bf{X}})$, $E[Y(1)-Y(0)|{\bf{U}}=0,{\bf{X}}]=E_{{\mathcal{Q}}}[Y(1)-Y(0)]-r$ and $E[Y(1)-Y(0)|{\bf{U}}=1,{\bf{X}}]=E_{{\mathcal{Q}}}[Y(1)-Y(0)]$ and the upper bound in (\ref{bound}) is attained for the same distribution of ${\bf{U}}|{\bf{X}}$ except that $E[Y(1)-Y(0)|{\bf{U}}=0,{\bf{X}}]=E_{{\mathcal{Q}}}[Y(1)-Y(0)]+r$.  Note that this distribution of ${\bf{U}}|{\bf{X}}$ corresponds to ${\bf{U}}=0$ being always takers or never takers, ${\bf{U}}=1$ being compliers.
\end{proof}

\begin{proof}[Proof of Proposition 4]
We have that
\begin{gather*}
\frac{E[E(Y|Z=1,{\bf{X}})-E(Y|Z=0,{\bf{X}})]}{E[P(D=1|Z=1,{\bf{X}})-P(D=1|Z=0,{\bf{X}})]}-PSIVWATE = \\
\frac{E[\{ E(Y|Z=1,{\bf{X}})-E(Y|Z=0,{\bf{X}})\}1\{ {\bf{X}},{\bf{U}}\in {\mathcal{A}}\}]}{E[P(D=1|Z=1,{\bf{X}})-P(D=1|Z=0,{\bf{X}})]} + \\
\frac{E[\{ E(Y|Z=1,{\bf{X}})-E(Y|Z=0,{\bf{X}})\}1\{{\bf{X}},{\bf{U}}\in {\mathcal{A}}^C\}]}{E[P(D=1|Z=1,{\bf{X}})-P(D=1|Z=0,{\bf{X}})]}-PSIVWATE = \\
PSIVWATE \frac{E[w({\bf{X}},{\bf{U}})1\{ {\bf{X}},{\bf{U}}\in {\mathcal{A}}\}]}{E[P(D=1|Z=1,{\bf{X}})-P(D=1|Z=0,{\bf{X}})]} + \\
NSIVWATE \frac{E[w({\bf{X}},{\bf{U}})1\{ {\bf{X}},{\bf{U}}\in {\mathcal{A}}^C\}]}{E[P(D=1|Z=1,{\bf{X}})-P(D=1|Z=0,{\bf{X}})]} - PSIVWATE = \\
-\lambda (NSIVWATE-PSIVWATE).
\end{gather*}
\end{proof}

\bibliographystyle{plainnat}
\bibliography{stochastic_monotonicity_refs}

\end{document}